\documentclass[useAMS,usenatbib]{mn2e}
\usepackage{amsmath,fleqn,graphicx,amssymb}
\usepackage{multirow}
\def\df{\textsc{df}}
\def\bolth{\mbox{\boldmath$\theta$}}
\def\bolOm{\mbox{\boldmath$\Omega$}}

\def\kpc{\,\mathrm{kpc}}
\def\kms{\,\mathrm{km\,s}^{-1}}

\def\bolJ{\mbox{\boldmath{$J$}}}
\def\bolx{\mbox{\boldmath{$x$}}}
\def\bolv{\mbox{\boldmath{$v$}}}
\def\vsol{\mbox{\boldmath{$v_\odot$}}}
\def\pc{\,\mathrm{pc}}

\def\d{\mathrm{d}}
\def\Rc{R_\mathrm{c}}

\title[Solar neighbourhood in angle coordinates]
{The Solar neighbourhood in angle coordinates: the Hyades moving group}

\author[P.~J.~McMillan]{
  Paul~J.~McMillan\thanks{E-mail: p.mcmillan1@physics.ox.ac.uk}\\
  Rudolf Peierls Centre for Theoretical Physics, 1 Keble Road,
  Oxford, OX1 3NP, UK
}

\begin{document}
\maketitle

\begin{abstract}
  I investigate the suggestion that the Hyades moving group in the 
  Solar neighbourhood is the result of a recent inner Lindblad resonance. 
  I use dynamical
  ``torus'' models of the Galaxy to understand the expected
  distribution of solar neighbourhood stars in angle coordinates for 
  phase-mixed models and models which include a resonant component. I
  show that attempts to find the signatures of resonances in angle
  coordinates are strongly influenced by selection effects, including rather
  subtle effects associated with the relationship between action and
  angle for stars at a given point. These
  effects mean that one can not use simple tests to determine whether 
  substructures seen
  in the Solar neighbourhood are associated with any given resonance.
\end{abstract}


\section{Introduction} \label{sec:intro} 
Since \cite{WD98} used proper
motions and parallaxes obtained by the Hipparcos satellite
\citep{Hipparcos} to investigate the kinematics of the Solar
neighbourhood, it has been clear that the local distribution function
(\df) is far from smooth. In particular, the distribution of stars in
the $U,V$ plane\footnote{Throughout this paper, in the Solar
  neighbourhood, velocities with respect to the Sun are described in
  terms of a component towards the Galactic Centre ($U$), a component
  in the direction of Galactic rotation ($V$), and a component
  perpendicular to the Galactic plane towards the north Galactic pole
  ($W$). Velocities with respect to the local standard of rest are
  described in terms of components in the same directions as $U$, $V$
  and $W$ which are given the symbols $v_x$, $v_y$ and $v_z$
  respectively} is dominated by a number of streams, all of which are
thought to be dynamical in origin \citep[e.g.][]{Faea05}.

Recently \citet[henceforth S10]{Se10} argued that part of this
substructure could be explained by a recent inner Lindblad resonance (ILR),
a conclusion he supported with reference to the distribution in angle
coordinates of stars observed by the Geneva-Copenhagen survey
\citep*[GCS:][]{GCS09}. This conclusion was supported by
\cite*{HaSePr11} who looked at stars in the Solar neighbourhood
observed by the Radial Velocity Experiment
\citep[RAVE:][]{RAVE1_short} and the Sloan Digital Sky Survey
\citep[SDSS:][]{SDSS7_short}.

In this paper I reexamine S10's conclusion that the distribution of
stars in the Solar neighbourhood show signs of an inner Lindblad
resonance (ILR). I compare the GCS sample of stars in the Solar
neighbourhood to a phase-mixed dynamical model. This allows me to
separate selection effects from genuine substructure in the local \df
, and to develop some understanding of the impact of selection effects
on the observed properties of any substructure that is found in the
local \df . I then explore the impact of selection effects on simple models
of an ILR or an outer Lindblad resonance (OLR).

In Section~\ref{sec:AAcoords} I discuss angle-action coordinates, how they 
might be used to determine the dynamical origin of observed kinematic 
substructure, and
their relationship to kinematics in the Solar neighbourhood. In
Section~\ref{sec:num} I give numerical details of the assumptions made
and the phase-mixed dynamical model considered. In Section~\ref{sec:smooth} I
explore the appearance in angle coordinates of the solar neighbourhood
in a phase-mixed model, which I then use in Section~\ref{sec:GCS} to
interpret the distribution in angle coordinates of stars observed by
the GCS. Section~\ref{sec:resmod} discusses simple models which include 
a resonant component, and uses them to better understand the GCS data.

\section{Angle-action coordinates}\label{sec:AAcoords}

Three actions $J_i$ and three conjugate angle coordinates $\theta_i$
provide canonical coordinates for stars orbiting in the
gravitational potential of the Galaxy. For a particle on any orbit the
actions are conserved quantities and the angles increase linearly with
time, $\theta_i(t) = \theta_i(0)+\Omega_it$, where $\Omega_i$ is a 
frequency. This means that $\bolJ$ can be thought of as labeling an orbit, and 
$\bolth$ as describing a point on that orbit. The usual phase space coordinates
$\bolx,\bolv$ are $2\pi$-periodic in each angle coordinate $\theta_i$.
For a phase-mixed \df\ the distribution of stars is uniform in angle
for any given range of actions, and therefore the \df \ $f=f(\bolJ)$.

The usage of angle-action coordinates has been somewhat limited by the
fact that the relationship between $\bolx,\bolv$ and $\bolth,\bolJ$ is
only available analytically for a very limited set of gravitational
potentials. I use the ``torus-fitting'' method \citep[e.g.][]{PJMJJB08} 
to find angle-action coordinates for stars with known velocities, and 
to construct a
phase-mixed model. S10 found values for the angle-action coordinates in 
the plane using the approximation that the in-plane and vertical 
components of motion 
could be decoupled, and then integrating orbits in the plane. The torus
method requires no such approximation, but since the 
stars being considered do not generally move far from the plane of the 
Galaxy, these approaches produce very similar results.

It was shown by S10 that stars trapped or scattered at a resonance
with some perturbation in the potential will, at any given time,
satisfy the relation
\begin{equation}\label{eq:res}
  l\theta_r+m\theta_\phi \simeq const ,
\end{equation}
where the perturbation has $m$-fold rotational symmetry, and $l$ is an
integer or $l=\pm \frac{1}{2}$ for ultra-harmonic resonances. $l=-1$
corresponds to an ILR, $l=1$ corresponds to an
OLR, and $l=0$ at corotation.  The value of the constant in
eq.~\ref{eq:res} can take any value, and varies linearly with time. This 
condition is in addition to the usual condition for resonance
\begin{equation}\label{eq:res_om}
  l\Omega_r+m\Omega_\phi = const = m\Omega_p ,
\end{equation}
where $\Omega_p$ is the pattern speed of the perturber, and $l$ \& $m$ have 
the same meaning as before. This condition on 
frequency can also be thought of as being a condition on action, since 
$\bolOm=\bolOm(\bolJ)$

The zero-points for the angles can be defined arbitrarily (provide the
same convention is applied for all orbits), and for clarity I follow
the conventions used by S10. This means that each component of $\theta_i$ 
lies in the range $[-\pi,\pi]$. The Galactocentric coordinates are aligned
such that the Sun is at a position in real
space with Galactocentric coordinate $\phi=0$.  I take $\theta_r=0$ at
apocentre, and therefore $\theta_r=\pm\pi$ at pericentre. I define
the zero point of $\theta_\phi$ such that at apocentre $\theta_\phi=\phi$. 
Note that for small $J_r,J_z$, where it is appropriate to 
use the epicycle approximation, the value of $\theta_\phi$
corresponds to the position (in $\phi$) of the guiding centre.

\begin{figure}
  \centerline{\resizebox{\hsize}{!}{\includegraphics[angle=270]{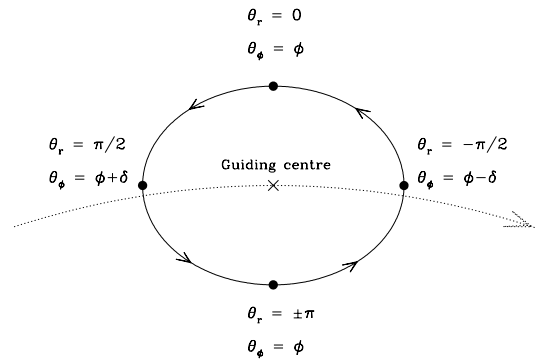}}}
  \caption{
    Schematic diagram illustrating the relationship between $\phi$ and 
    $\theta_\phi$ as a function of $\theta_r$ for low eccentricity orbits, 
    for which the epicycle approximation applies. 
    The \emph{dotted} line and arrow 
    shows the motion of the guiding centre (shown as a cross). 
    The \emph{solid} line and arrows show the motion of the star on an epicycle
    around the guiding centre. The maximum difference 
    between $\phi$ and $\theta_\phi$ is $\delta$ which can be found (under 
    the epicycle approximation) using eq.~\ref{eq:delta}. Note that the 
    epicycle approximation is used purely for illustrative purposes in this 
    paper, and is not used to find the relationship between $\bolx,\bolv$
    and $\bolth,\bolJ$. 
\label{fig:scheme}
}
\end{figure}

\begin{figure*}
  \centerline{\hfil\resizebox{82mm}{!}{\includegraphics{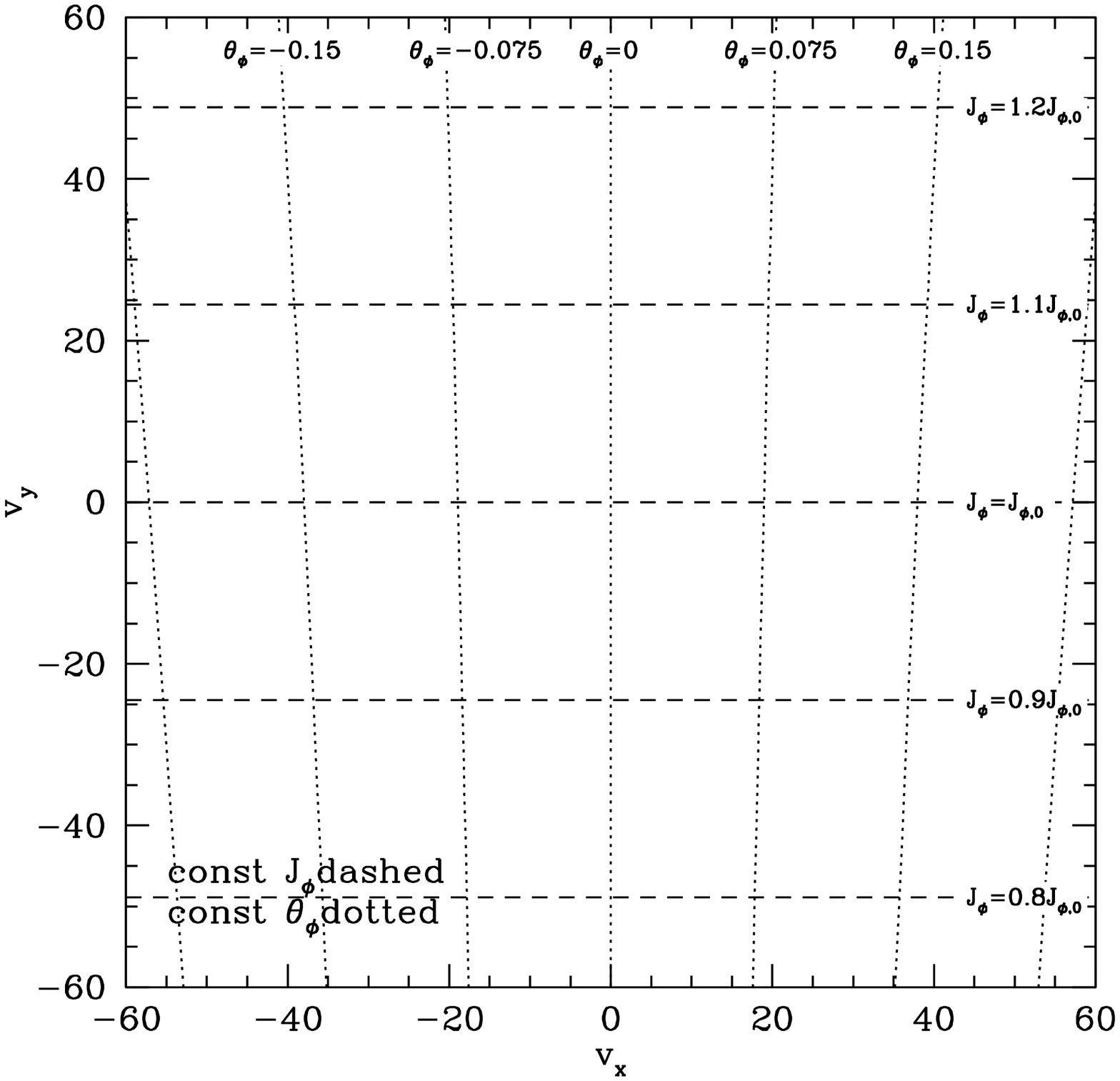}}
    \hspace{4mm} \resizebox{82mm}{!}{\includegraphics{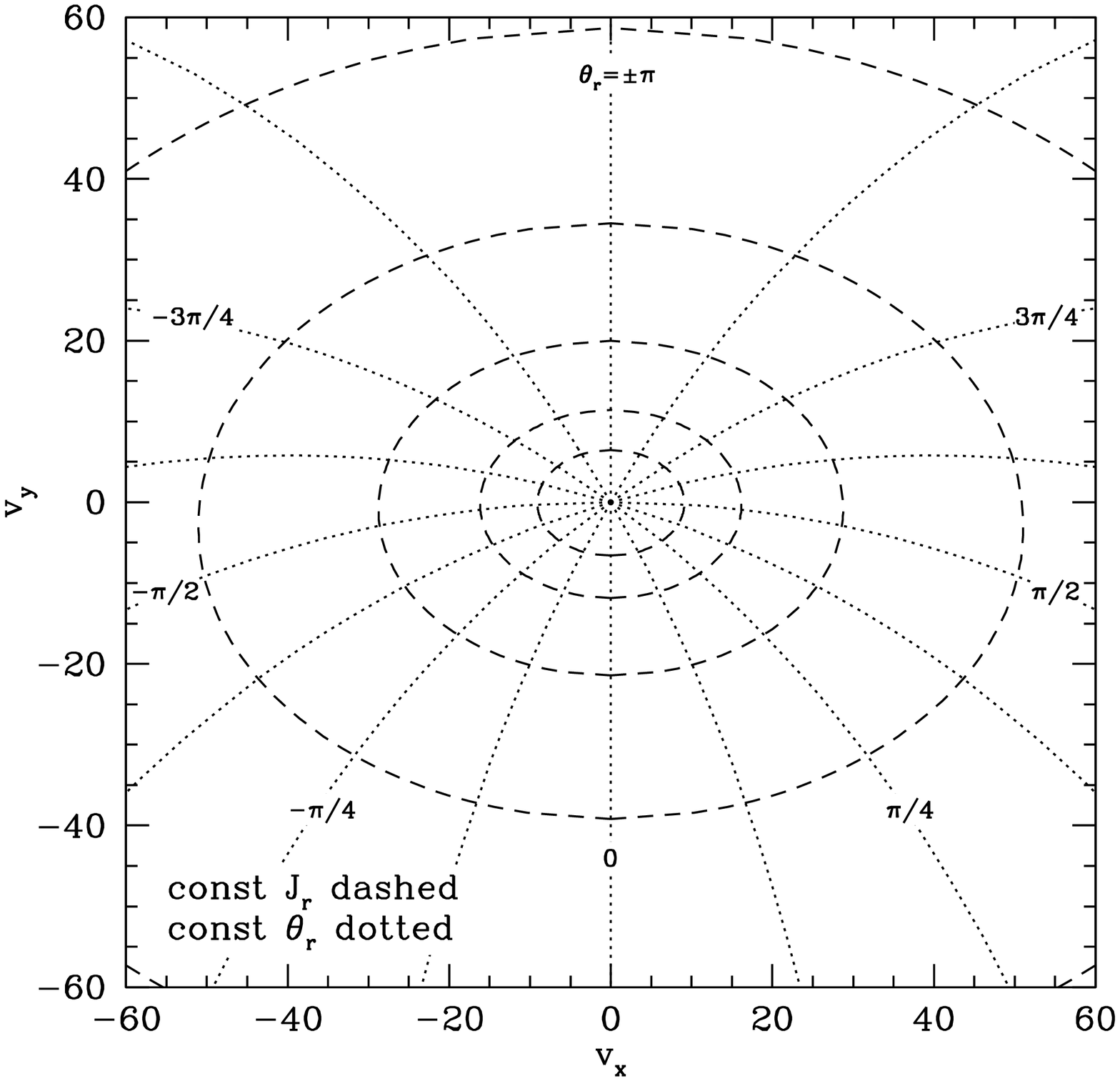}}}
  
  \caption{The $v_x,v_y$ plane, with lines of constant 
    $\theta_\phi$ or $J_\phi$ (dotted
    and dashed respectively, left panel) or lines of constant $\theta_r$ or 
    $J_r$ (again dotted and dashed respectively, right panel) plotted, for a 
    star at the Sun's position with $J_z=0$. 
    The lines of constant $\theta_\phi$ and $J_\phi$ 
    are linearly spaced in each case and labelled on the plot (where 
    $J_{\phi,0}=2080\kpc\kms$ is the angular momentum of a circular orbit 
    at the Solar 
    radius) -- they form a nearly-Cartesian coordinate system for the 
    $v_x,v_y$ plane. The lines of constant $\theta_r$ are shown at intervals 
    of $\pi/8$, with $\theta_r$ increasing as you move anti-clockwise on 
    the plot. The lines of constant $J_r$ are logarithmically spaced in 
    $J_r$, with the innermost one at $1\kpc\kms$ and the value of $J_r$
    increasing by a factor of $\sqrt{10}$ at a time, 
    with the outermost visible line corresponding to $J_r=100\kpc\kms$.  
    $J_r$ and $\theta_r$ form a nearly-polar coordinate set in the 
    $v_x,v_y$ plane, with $J_r$ as a
    ``radius'' and $\theta_r$ as a ``polar angle''.    
    \label{fig:UVaxes}
  }
\end{figure*}

Figure~\ref{fig:scheme} is a schematic diagram illustrating the
relationship between $\theta_r$ and $\theta_\phi$ using the example of
an orbit in the plane of the Galaxy with small eccentricity, where use
of the epicycle approximation is appropriate. The star moves on an
epicycle about the guiding centre, which is itself moving at a
constant angular velocity. The radius of the orbit of the guiding
centre about the galactic centre is determined by the angular 
momentum of the orbit. The size of the epicycle (i.e. $a$, the maximum radial 
excursion of the orbit away from the guiding centre and $\delta$, the maximum 
value of $|\phi-\theta_\phi|$) can
be found under the epicycle approximation
\begin{equation}\label{eq:a_epi}
a\approx\sqrt{2J_r/\kappa}
\end{equation}
and
\begin{equation}\label{eq:delta}
\delta\approx\frac{2\Omega_\mathrm{c}}{\kappa R_g}\,\times\sqrt{2J_r/\kappa},
\end{equation}
where $R_g$ is the radius of the guiding centre, $\Omega_\mathrm{c}(J_\phi)$ 
is the circular frequency,
and $\kappa(J_\phi)$ is the radial epicycle frequency \citep[e.g.][]{GDII}.
For given value of $\bolJ$ the value of
$\theta_r$ sets the relationship between $\theta_\phi$ and $\phi$, and
the orbit only goes through a given point (e.g. the Sun's position) at
two values of $\bolth$ (related through $\theta_{r,1}=-\theta_{r,2}$,
$\theta_{\phi,1}-\phi=\phi-\theta_{\phi,2}$).

In Figure~\ref{fig:UVaxes} I show how values of $J_r$, $J_\phi$,
$\theta_r$ or $\theta_\phi$ correspond to lines in the $v_x,v_y$-plane
for stars that are exactly at the Sun's position, and for which we
can ignore motion out of the plane (i.e.~$J_z=0$). As that plot shows,
one can think of $J_\phi$ and $\theta_\phi$ as providing
near-Cartesian coordinate axes in the $v_x,v_y$-plane, and $J_r$ and
$\theta_r$ as providing near-polar coordinate axes in the
$v_x,v_y$-plane. Indeed specifying any two values from the set
$(J_r,J_\phi,\theta_r,\theta_\phi)$ defines a single position in the
$v_x,v_y$-plane for a star at the Sun's position -- except if the
values are $J_r$ and anything other than $\theta_r$, in which case it
can describe zero, one or two possible positions in the
$v_x,v_y$-plane.

The stars considered in this paper are those found within $200\pc$ of
the Sun, but none actually at the Sun's position, so the relationships
between $v_x,v_y$ and $\bolJ,\bolth$ shown in Figure~\ref{fig:UVaxes}
are only approximate (even ignoring the motion out of the plane).  A
star being at Galactocentric $\phi=\phi_\ast\neq0$ affects the value
of $\theta_\phi$, shifting it by $\phi_\ast$ for a given $v_x,v_y$,
with the greatest value of $\phi_\ast$ possible for this sample being
$\sim0.02$. A star being at $R\neq R_0$ affects the value of
$J_\phi$ for a given $v_x,v_y$, but only by maximum of $\sim0.02
J_{\phi,0}$ in this sample (with $J_{\phi,0}$ the angular momentum of a 
circular orbit at $R_0$).  The values of $J_r$ or $\theta_r$ for a
given $v_x,v_y$ are not significantly affected by the star's position
within the solar neighbourhood, primarily because the circular
velocity (and thus the velocity of a $J_r=0$ orbit) barely changes
across the volume considered for any reasonable Galactic potential.

The rest of this paper focuses on the distribution of stars in $\theta_r$
and $\theta_\phi$. I have examined the distribution of stars in 
$\theta_z$ and, apart from the expected tendency for stars to be at 
values of $\theta_z$ that place them near the Galactic plane (a selection 
effect), it shows no interesting features. This is entirely in keeping 
with the absence of kinematic substructure in $v_z$ noted by \cite{WD98}, 
and S10's brief discussion of the $J_z$ distribution.

\section{Numerical details} \label{sec:num}

In all cases I use the ``convenient'' model Galactic potential given
by \cite{PJM11:mass}. This model consists of a bulge component, thin
and thick exponential discs, and a \cite*{NFW96} halo.  This sets the
solar radius $R_0=8.5\kpc$ and the circular velocity at the Sun (the
local standard of rest) $v_0=244.5\kms$. I have explored the effect of using
alternative Galactic potentials (including a logarithmic potential of
the kind used by S10), and it does not significantly alter the main results.

I assume that the velocity of the Sun with respect to the local
standard of rest is the best-fitting value found by \cite*{SBD10}
\begin{equation}\label{eq:vsol}
  \vsol  =   (U_\odot,V_\odot,W_\odot) =  (11.1,12.24,7.25)\kms .
\end{equation}

The phase-mixed model \df\ I compare to the real data is very similar
to that described in \cite{JJBPJM11:dyn}, but with altered disc
scale-lengths (to reflect the scale-lengths of the discs that produce
the potential). The thin and thick discs are modelled as having
``quasi-isothermal'' \df s, which is to say that they are of the form
\begin{equation} \label{eq:totalDF}
  f(J_r,J_\phi,J_z)=f_{\sigma_r}(J_r,J_\phi)\times \frac{\nu_z}
  {2\pi\sigma_z^2}\,\mathrm{e}^{-\nu_z J_z/\sigma_z^2},
\end{equation}
where
\begin{equation} \label{eq:planeDF}
  f_{\sigma_r}(J_r,J_\phi)\equiv\frac{\Omega_\mathrm{c}\Sigma}{\pi\sigma_r^2\kappa}\bigg|_{\Rc}
  [1+\tanh(J_\phi/L_0)]\mathrm{e}^{-\kappa J_r/\sigma_r^2}.
\end{equation}
Here $\nu(J_\phi)$  is the vertical epicycle frequency and 
$\Sigma(J_\phi)=\Sigma_0\mathrm{e}^{-(\Rc-R_0)/R_\d}$ is the
(approximate) radial surface-density profile, where $\Rc(J_\phi)$ is
the radius of the circular orbit with angular momentum $J_\phi$. The
factor $1+\tanh(J_\phi/L_0)$ in equation (\ref{eq:planeDF}) is there
to effectively eliminate stars on counter-rotating orbits and the
value of $L_0$ is unimportant provided it is small compared to the
angular momentum of the Sun. In equations (\ref{eq:totalDF}) and
(\ref{eq:planeDF}) the functions $\sigma_z(J_\phi)$ and
$\sigma_r(J_\phi)$ control the vertical and radial velocity
dispersions, and I set
\begin{eqnarray}\label{eq:sigmas}
  \sigma_r(J_\phi)&=&\sigma_{r0}\,\mathrm{e}^{q(R_0-\Rc)/R_\d}\nonumber\\
  \sigma_z(J_\phi)&=&\sigma_{z0}\,\mathrm{e}^{q(R_0-\Rc)/R_\d},
\end{eqnarray}
where $q=0.45$ and $\sigma_{r0}$ and $\sigma_{z0}$ are approximately
equal to the radial and vertical velocity dispersions at the Sun. I
take the \df\ of the entire disc to be the sum of a \df\ of the form
(\ref{eq:totalDF}) for the thin disc, and a similar \df\ for the thick
disc, the normalisations being chosen so that at the Sun the surface
density of thick-disc stars is 23 per cent of the total stellar
surface density. Table \ref{tab:df} lists the parameters of each
component of the \df.

The physical properties of the model are determined by both the \df \ and
the Galactic potential and, as in \cite{JJBPJM11:dyn}, the \df \ does 
not self-consistently reproduce
the potential. To produce a self-consistent dynamical model one has to 
specify a \df \ for the dark matter (which makes a substantial contribution
to the potential), and distinguish carefully between the masses and 
luminosities of the stars. This lies well beyond the scope (or needs) of this 
study.

\begin{table}
  \caption{Parameters of the \df.}\label{tab:df}
  \begin{center}
    \begin{tabular}{l|cccc}
      Disc & $R_\d/\hbox{kpc}$ & $\sigma_{r0}/\!\kms$ & $\sigma_{z0}/\!\kms$ &
      $L_0/\!\kpc\kms$\\
      \hline
      Thin & 3.0 & 27 & 20 & 10\\
      Thick& 3.5 & 48 & 44 & 10\\
    \end{tabular}
  \end{center}
\end{table}

Samples of stars are produced by sampling the
\df\ in $\bolJ$, then sampling evenly in
$\bolth$ in the range $[-\pi,\pi]$. This fully determines the 
position and velocity of the stars, and only those 
that lie within $200\pc$ of the Sun become part of the sample. 
This is the same procedure that was 
used in \cite{JJBPJM11:dyn} and, as in that paper, $200,000$ values 
of $\bolJ$ are sampled from the \df \ 
(all of which correspond to orbits that enter the solar neighbourhood) from 
which $1,000,000$ stars in the solar neighbourhood are sampled.

\begin{figure}
  \centerline{\hfil\resizebox{45mm}{!}{\includegraphics{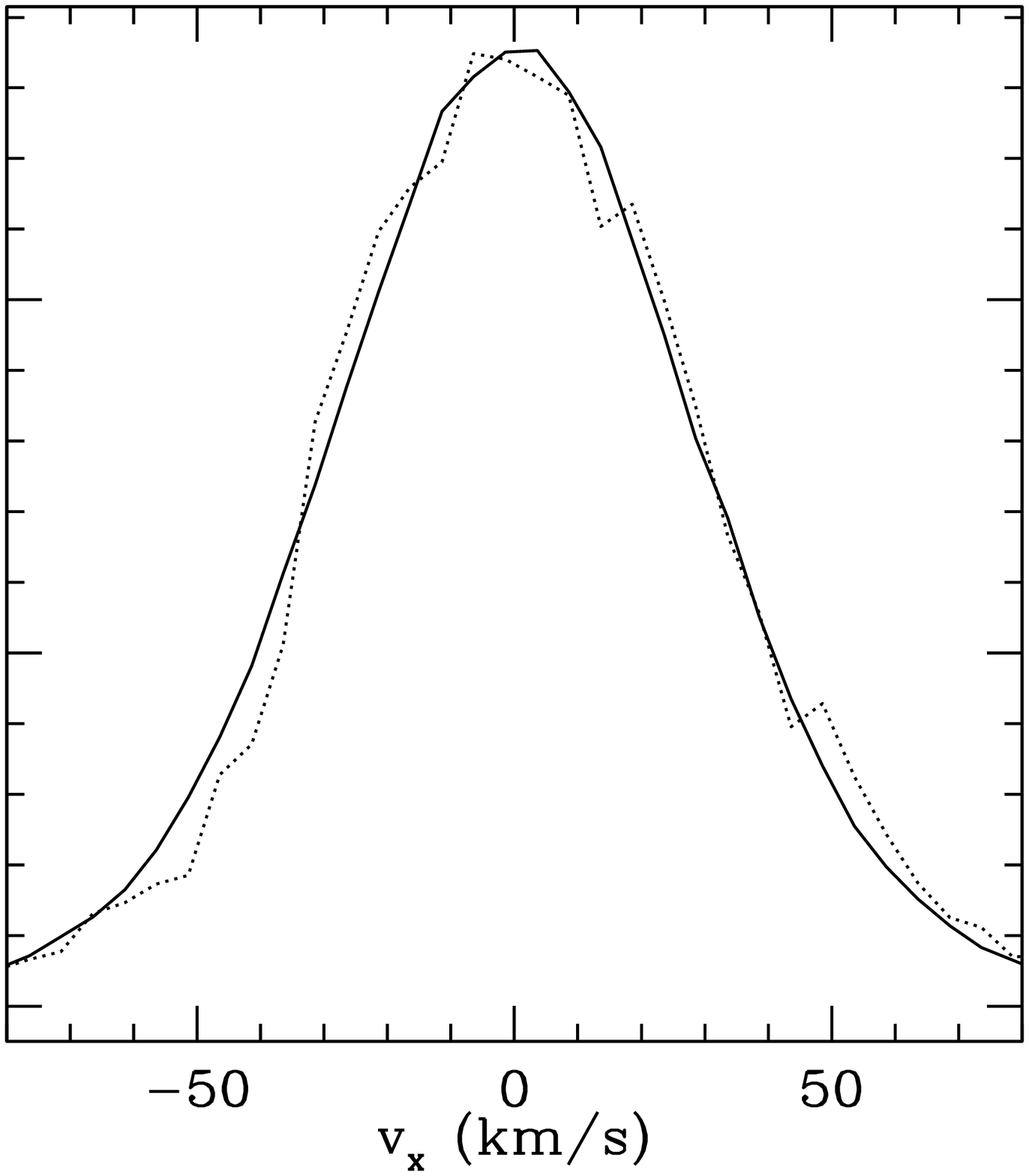}}
\hspace{-3mm}
   \resizebox{45mm}{!}{\includegraphics{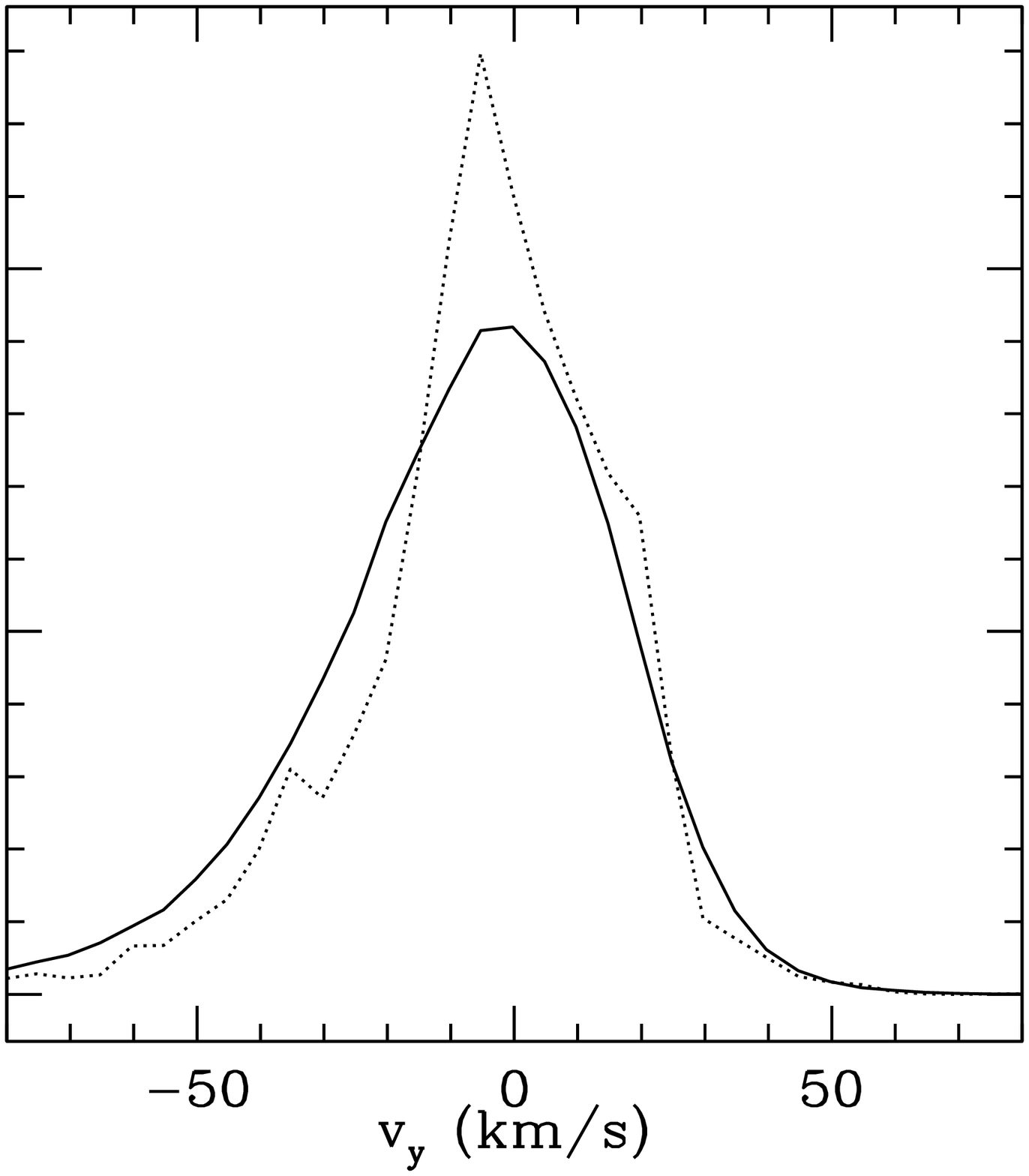}}}  
  \caption{ 
    Distribution in $v_x$ (\emph{left}) and $v_y$ (\emph{right}) 
    of stars observed by the GCS 
    (assuming $\vsol$ is as given in eq.~\ref{eq:vsol}, \emph{dotted} line) 
    and taken from the solar neighbourhood in the
    phase-mixed model (\emph{solid} line).
    \label{fig:UVhist}
  }
\end{figure}

In Figure~\ref{fig:UVhist} I plot the distribution of stars in $v_x$ and 
in $v_y$ from the GCS (assuming the value of $\vsol$ given in 
eq.~\ref{eq:vsol}) and from the phase-mixed \df . 
This simple \df \ does not entirely fit the observed 
data, in particular the GCS data has a much higher peak in its $v_y$ 
distribution 
than the model data. However the $v_x$ distributions are very similar and the 
model $v_y$-distribution is appropriately skew. Since this model is 
only being used 
as a guide, rather than as an attempt to reproduce the observed 
distributions, this is sufficient and I do not attempt to fit the 
observed distribution more precisely with \df s that are more 
complicated functions of the actions~\citep[e.g.][]{JJB10}.

\section{Appearance of a phase-mixed component in angle
  coordinates}\label{sec:smooth}

\begin{figure}
  \centerline{\resizebox{\hsize}{!}{\includegraphics{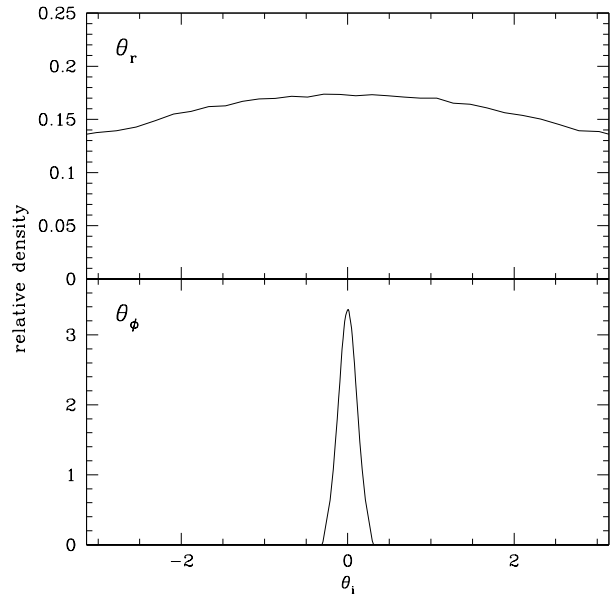}}}
  \caption{Distribution in $\theta_r$ (\emph{upper}) and $\theta_\phi$ 
    (\emph{lower}) of stars in the solar neighbourhood 
    (radius of $200\pc$) taken from the \df\ given in 
    equation~\ref{eq:totalDF}. These histograms, and all others in this paper 
    showing the relative density of stars as a function of the angles,
    are normalised such that the total area under the histogram is unity, and 
    are found by binning stars with bin width $\sim0.1$ radians.  
\label{fig:mod_0}
}
\end{figure}

\begin{figure}
  \centerline{\resizebox{\hsize}{!}{\includegraphics[angle=270]{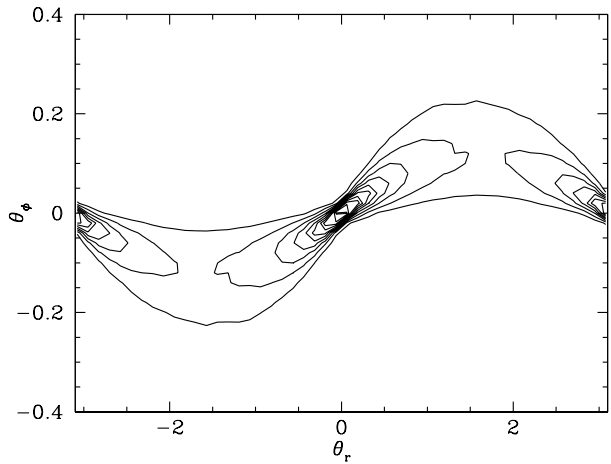}}}
  \caption{Contour plot showing the density in the 
    $\theta_r,\theta_\phi$ plane of stars in the solar neighbourhood
    (radius of $200\pc$) taken from the \df\ given in 
    equation~\ref{eq:totalDF}. Contours are spaced linearly in density. Note 
    that the regions near $\theta_r=0,\pm\pi$ have a high density because all
    stars near those points have $\phi\approx\theta_\phi$, and so stars
    in the solar neighbourhood have values of $\theta_\phi$ 
    that lie in a very small range (which does not vary with $J_r$).
\label{fig:mod_0_cont}
}
\end{figure}

While, for a phase-mixed population, the distribution of stars in
angle is uniform over any given range in action, it does not
generally follow that the distribution is uniform over a given volume
in real space. The distribution in angle of stars found within the
Solar neighbourhood in a phase-mixed model is significantly
non-uniform. In Figure~\ref{fig:mod_0} I show the distribution in
$\theta_r$ and in $\theta_\phi$ of stars in the solar neighbourhood,
taken from the phase-mixed model.

The narrow distribution in $\theta_\phi$ reflects the fact that the
stars are taken from a very narrow range in $\phi$, and that there is a close 
relationship between $\phi$ and $\theta_\phi$. The average value 
of $\delta$ (approximated using eq.~\ref{eq:delta}) for stars in the 
survey volume is $\sim0.16$.

\begin{figure*}
  \centerline{\resizebox{\hsize}{!}{\includegraphics[angle=270]{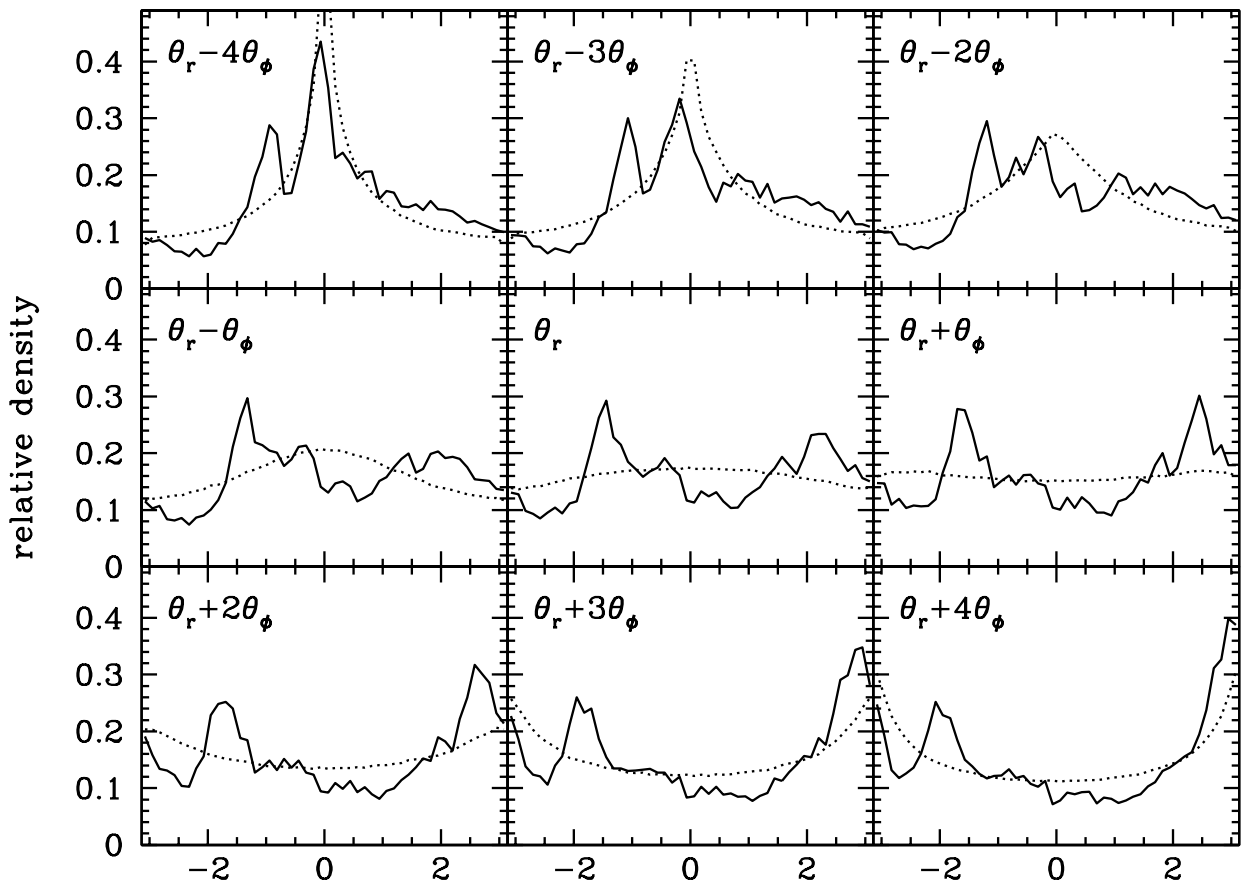}}}
  \caption{Plot showing the relative density of stars in the solar 
    neighbourhood taken from my phase-mixed model (\emph{dotted} line) and the GCS data 
    (\emph{solid} line) for the linear combinations of angles $\theta_r+n\theta_\phi$ 
    for various values of $n$. The dotted line illustrates the effects of 
    selection effects in each plot, so peaks in the solid line that are not 
    seen in the dotted line are genuine substructure. ILRs 
    are expected to produce peaks for $n<0$ and OLRs to produce peaks for $n>0$.
    The main features are (a) 
    the peak visible in all plots that moves gradually with increasing $n$ 
    from $\theta_r-4\theta_\phi\sim-1$ to $\theta_r+4\theta_\phi\sim-2$; and 
    (b) the peak that is only clearly visible for $n\geq0$, moving with 
    increasing $n$ from $\theta_r\sim2$ to $\theta_r+4\theta_\phi\sim3$. The 
    movement and (to at least some extent) sharpening or broadening 
    of the peaks is due to the selection effects seen in 
    Figure~\ref{fig:mod_0_cont}.
    \label{fig:meat}
  }

\end{figure*}

The relative density of stars in $\theta_r$ is at a maximum around $0$
(i.e.  apocentre) and a minimum around $\pm\pi$
(pericentre). The stars at apocentre have guiding radii smaller than 
their current radius, and those at pericentre have guiding radii larger 
than their current radius. The excess of stars at apocentre is due to 
the fact that the
density of stars, and their velocity dispersion, decreases with
increasing radius, so more stars visit the Solar neighbourhood from 
guiding radii smaller than $R_0$ than from guiding radii larger than $R_0$.
Since stars at apocentre are lagging circular rotation, while those 
at pericentre are leading it, this non-uniformity in the $\theta_r$ 
distribution is directly related to asymmetric drift, and the skew
distribution in $v_y$ seen in Figure~\ref{fig:UVhist} \citep[e.g.][]{GDII}. 
This behaviour is different from that suggested by S10 who
incorrectly claimed that one should expect low relative
density of stars around both $\theta_r=\pm\pi$ and $\theta_r=0$.


In Figure~\ref{fig:mod_0_cont} I show density contours for the
values of $\theta_r$ and $\theta_\phi$ taken from the phase-mixed
model. There is a clear relationship between the two values, with
$\theta_\phi\gtrsim0$ for $\theta_r>0$, and $\theta_\phi\lesssim0$ for
$\theta_r<0$. There are extrema in $\theta_\phi$ at
$\theta_r=\pm\pi/2$, and $\theta_\phi\approx0$ for $\theta_r=0$ \&
$\pm\pi$.  This is again due to the fact I am selecting stars from a
very narrow range in $\phi$, and because of the relationship between
$\theta_r$ and $\theta_\phi$ for a given orbit at a given point
illustrated in Figures~\ref{fig:scheme} and \ref{fig:UVaxes}.


\section{Geneva Copenhagen survey data} \label{sec:GCS}

I now compare the phase-mixed model analysed in
Section~\ref{sec:smooth} to the stars in the Solar neighbourhood
observed by the GCS. The GCS data are taken from the table produced
by~\cite{GCS09}, and I follow S10 in restricting the analysis to stars
that have full 6D phase-space coordinates quoted, are at distances
$\leq200\pc$ from the Sun, and are not directly associated with the
Hyades cluster. The distributions in angle space that I find are very
similar to those found by S10, and the small differences are due to
the different choices of Galactic potential and the use of torus-fitting
as opposed to integrating orbits in the plane.

In Figure~\ref{fig:meat} I plot the distributions in
$\theta_r+n\theta_\phi$ of the GCS stars (solid line) and those taken
from the phase-mixed model (dotted line) for a series of integers
$n$. Naively one would simply expect an OLR to 
produce a peak
for some value of $n>0$, and an ILR to produce a
peak for some value of $n<0$ (with the perturber being $|n|$-fold
symmetric). The relationship between $\theta_r$ and $\theta_\phi$ due
to selection effects produces a peak around $0$ in plots with $n<0$
and around $\pm\pi$ for $n>0$.  \footnote{The convention I use for
  these graphs means that for $n<0$ they are mirror images of the
  comparable figures in S10. This choice is made so that the features
  in the various plots can easily be ``followed'' from one to another
  as $n$ increases.}

There are two main features in the angle distribution of the GCS data
(besides those due to selection effects), one peak that lies at
$\theta_r\sim-1.5$, and one that lies at $\theta_r\sim2.2$. It is also
noticeable that the peak at $0$ which appears in the model data for
$n<0$ is offset to slightly lower values in the GCS data -- this can
be associated with the small peak in the $\theta_r$ distribution at
$\theta_r\sim-0.4$.

\begin{figure}
  \centerline{\resizebox{\hsize}{!}{\includegraphics{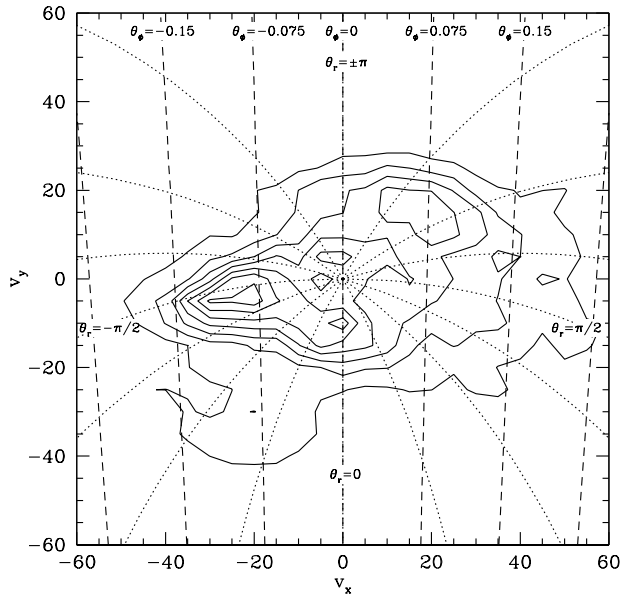}}}
  \caption{Distribution of GCS stars in the $v_x,v_y$-plane (\emph{solid 
      contours}), 
    overlaid on lines of constant $\theta_r$ (\emph{dotted}) and $theta_\phi$ 
    (\emph{dashed}) for stars at the Sun's position with $J_z=0$, which are 
    shown as a guide. This makes it 
    easy to associate features seen in Figure~\ref{fig:meat} with the familiar 
    Solar neighbourhood moving groups \citep[e.g.][]{WD98}. The 
    Hyades moving group (around $v_x=-25\kms$, $v_y=-5\kms$) is 
    clearly associated with the peak at $\theta_r\sim-1.5$. The Sirius moving 
    group (around $v_x=20\kms$, $v_y=15\kms$) is associated with the peak 
    at $\theta_r\sim2.2$. The Pleiades moving group (around $v_x=-2\kms$, 
    $v_y=-10\kms$ is associated with the peak near $\theta_r=-0.4$.
\label{fig:UV_GCS}
}
\end{figure}

Figure~\ref{fig:UV_GCS} shows the distribution of stars taken from the
GCS data in the $v_x,v_y$-plane (assuming the Solar velocity relative
to the local standard of rest given in eq.~\ref{eq:vsol}) overlaid on
the lines of constant $\theta_r$ and of constant $\theta_\phi$ (at the
Sun's position) as shown in Figure~\ref{fig:UVaxes}. This makes it
easy to see which of the familiar features in the $v_x,v_y$-plane
correspond to which peaks in Figure~\ref{fig:meat}. The peak at
$\theta_r\sim-1.5$ is associated with the Hyades moving group, the
peak at $\theta_r\sim2.2$ is associated with the Sirius moving group
and the peak at $\theta_r\sim-0.4$ is associated with the Pleiades
moving group.


The peak at $\theta_r\sim-1.5$ clearly appears in all the plots shown
in Figure~\ref{fig:meat}, shifted to higher values for $n<0$ and to
lower values for $n>0$. This peak is made up of almost the same set of
stars in each plot. It is this feature that S10 
identified as the signature of an ILR. It is
clear from Figure~\ref{fig:meat} that these data are consistent with
the stars that make up the peak at $\theta_r\sim-1.5$ being associated
with any value of $n$ between $-4$ and $4$ (including non-integer values, 
and indeed for $|n|>4$
though these are not shown as selection effects create ever greater
distortions). While S10 shows plots that are almost identical to the
centre and bottom-left panels of Figure~\ref{fig:meat} (his fig. 4 upper 
panel and fig. 7 top panel, respectively), the peaks in
these plots are misidentified as being primarily due to selection 
effects that S10 incorrectly claimed should result in high relative densities
around $\theta_r=\pm\pi/2$.

The peak at $\theta_r\sim2.2$ becomes a more clearly defined (and
higher) peak for $n>0$ (shifted to higher values), and nearly
disappears for $n<0$. It is also noticeable that the feature around
$\theta_r\sim-1.5$ becomes more sharply peaked for $n<0$. This would
seem to imply that the peak at $\theta_r\sim2.2$ is associated with an
OLR, and the peak at $\theta_r\sim-1.5$ is
associated with an ILR, but this is not
necessarily the case, both because of the selection effects discussed
previously, and because of the other condition on resonant stars 
(eq.~\ref{eq:res_om}, see Section~\ref{sec:resmod}).

\begin{figure}
  \centerline{\resizebox{\hsize}{!}{\includegraphics[angle=270]{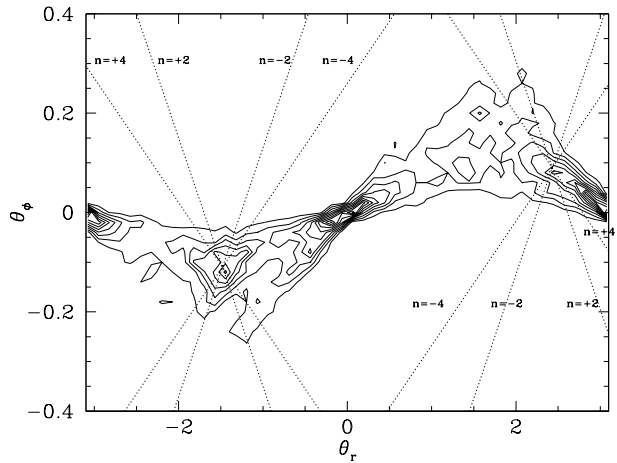}}}
  \caption{Contour plot of the density in the $\theta_r$, $\theta_\phi$ 
    plane of stars taken from the GCS data (\emph{solid} lines). 
    Contours are spaced linearly in density, at the 
    same densities as Figure~\ref{fig:mod_0_cont} (relative to the total 
    number of stars in each case). 
    The various plots in Figure~\ref{fig:meat} 
    can be thought of as being constructed from the density in this 
    plane by marginalising over lines of constant 
    $\theta_r+n\theta_\phi$. The \emph{dotted} lines show these lines for 
    $n=-4,-2,+2,+4$ passing through the main overdensities, to guide the eye.   
\label{fig:contGCS}
}
\end{figure}

In Figure~\ref{fig:contGCS} I show a contour plot of the density in
the $\theta_r$, $\theta_\phi$ plane of the selected stars from the GCS
catalogue.  The plots
in Figure~\ref{fig:meat} are found from the distribution in angle
space plotted in Figure~\ref{fig:contGCS} by marginalising it over
straight lines, the gradient of which are determined by $n$. In
Figure~\ref{fig:contGCS}, lines corresponding to a few different
values of $n$ are plotted, centred on the main overdensities, to guide
the eye.

\begin{figure}
  \centerline{\hfil\resizebox{45mm}{!}{\includegraphics{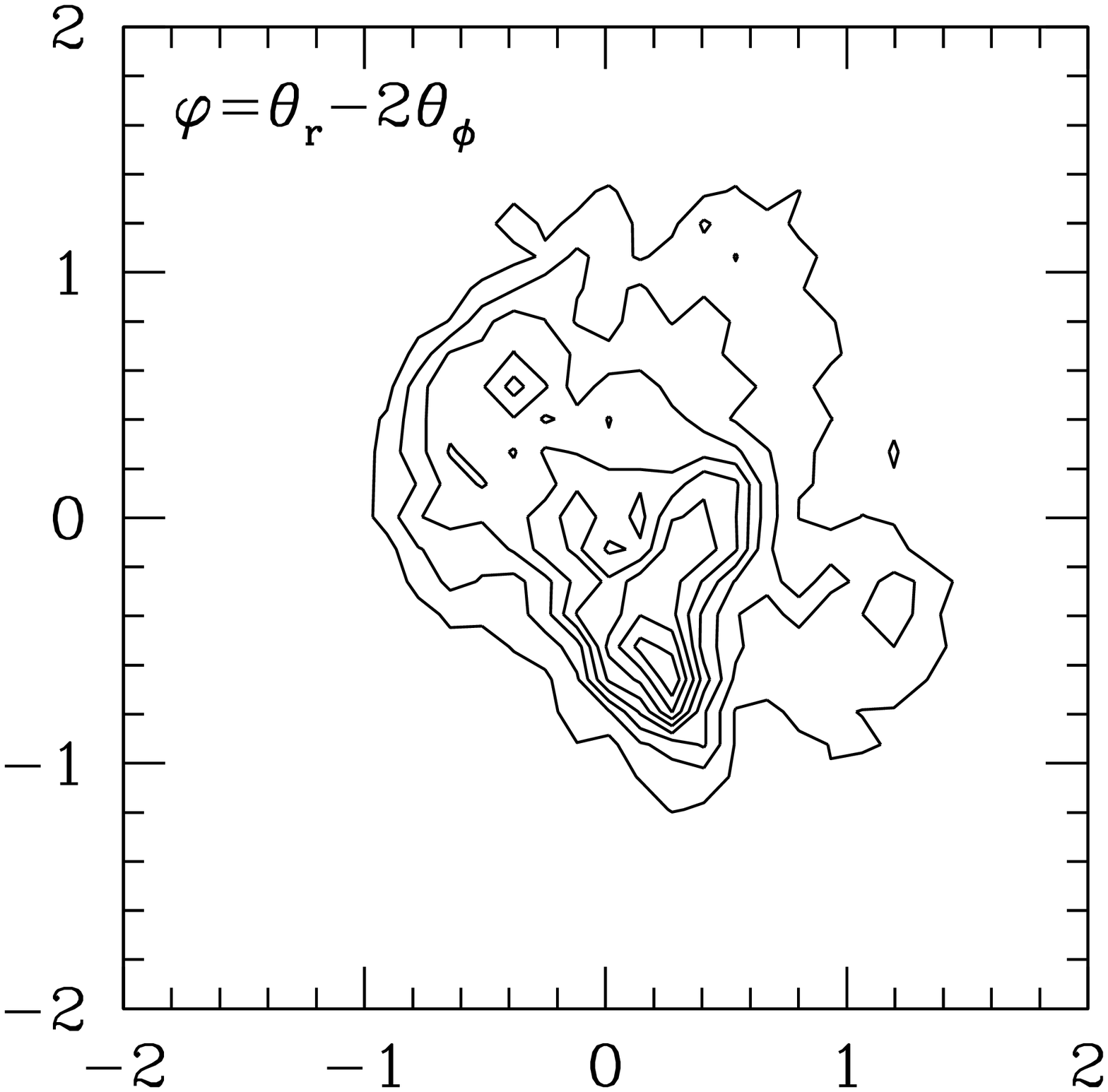}}
\hspace{-3mm}
   \resizebox{45mm}{!}{\includegraphics{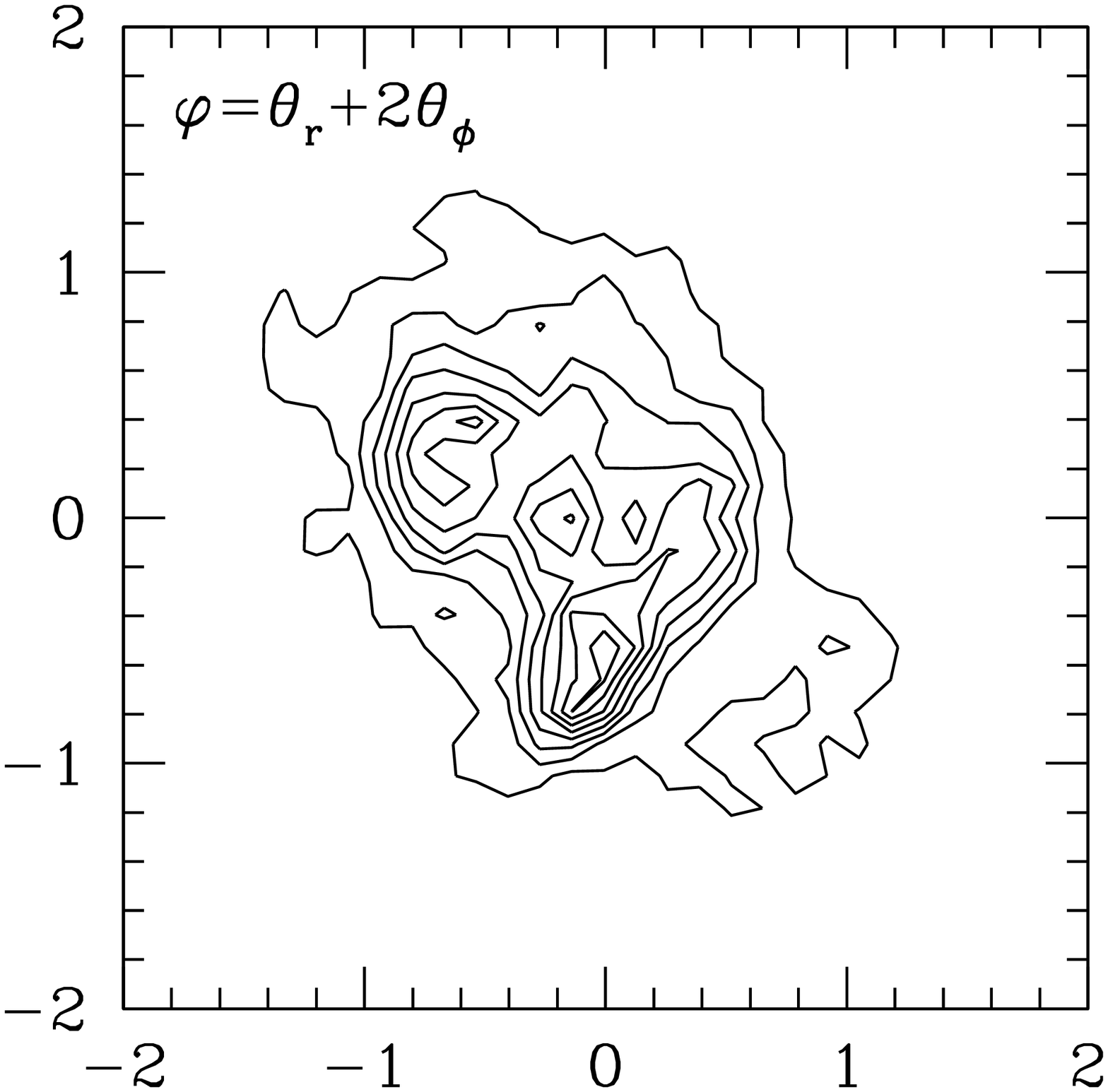}}}
  
  \caption{ 
    Contour plot showing the distribution of GCS stars in polar 
    coordinates with radial coordinate $a$ (eq.~\ref{eq:a_epi}) 
    the amplitude of radial motion under the 
    epicycle approximation, and polar coordinate 
    $\varphi=\theta_r+n\theta_\phi$ with $n=-2$ (\emph{left}) 
    or $n=2$ (\emph{right}). The overdensity corresponding to the 
    Hyades is found around $(x,y) = (0.3,-0.6)$ in the $n=-2$ plot and
    around  $(-0.05,-0.6)$ in the $n=2$ plot. A plot very similar to 
    the $n=-2$ one was shown by S10 as fig. 8. 
    \label{fig:pol_GCS}
  }
\end{figure}

The distribution in the  $\theta_r,\theta_\phi$ plane seen in 
Figure~\ref{fig:contGCS} shows the
same strong selection effects in $\bolth$ illustrated by the phase-mixed 
model. These selection effects have a very strong effect on the 
overdensities associated with the Pleiades and Sirius moving groups, which 
drives them towards the observed correlation between $\theta_r$ and 
$\theta_\phi$ in the two different cases. The Hyades overdensity 
(around $\theta_r=-1.5$) is a less straightforward case. It 
is approximately triangular in the $\theta_r,\theta_\phi$ plane, and lies 
either side of the expected minimum in $\theta_\phi$ at 
$\theta_r=-\pi/2$ (this minimum is seen in the phase-mixed model, 
Figure~\ref{fig:mod_0_cont}). 
The Hyades overdensity has increasing $\theta_\phi$ with increasing $\theta_r$ 
for $\theta_r>-\pi/2$, and with decreasing $\theta_r$ 
for $\theta_r<-\pi/2$ (Figure~\ref{fig:contGCS}), 
and these are the directions one would expect the
selection effects that affected the phase-mixed model 
to drive the observations. This selection effect is important, but
it is not clear that it is enough to explain the observed shape, and in 
Section~\ref{sec:resmod} I explore the distribution in angle space of models 
with a resonant component.

Figure~\ref{fig:pol_GCS} shows a contour plot of the GCS data 
distributed in polar coordinates with radial coordinate 
$a$ (the amplitude of radial motion under the epicycle approximation, 
eq.~\ref{eq:a_epi}) and polar coordinate 
$\varphi=\theta_r-2\theta_\phi$ (left) or $\varphi=\theta_r+2\theta_\phi$ 
(right). These are comparable to the upper panel of S10's fig. 8. In 
both cases one can see an overdensity at particular values of 
the polar coordinate, found (unsurprisingly) at the values of 
$\varphi$ one would anticipate given Figure~\ref{fig:meat}. S10 
only showed a version of the plot with $\varphi=\theta_r-2\theta_\phi$, 
finding that the plot for $\varphi=\theta_r+2\theta_\phi$ (i.e. the 
plot that would suggest an OLR) argued against the suggestion that 
the Hyades corresponded to an OLR, primarily because the overdensity 
in the $\varphi=\theta_r-2\theta_\phi$  plot appears to lie in a radial line 
out to larger $a$ in the figure, whereas in the 
$\varphi=\theta_r+2\theta_\phi$ plot it does not (Sellwood priv. comm.). 
This is not a particularly strong effect, but can be seen in 
Figure~\ref{fig:pol_GCS}, and understanding 
it requires moving beyond a phase-mixed model.

\section{Appearance of an OLR or ILR in angle  
coordinates} \label{sec:resmod}

\begin{figure}
  \centerline{\resizebox{\hsize}{!}{\includegraphics{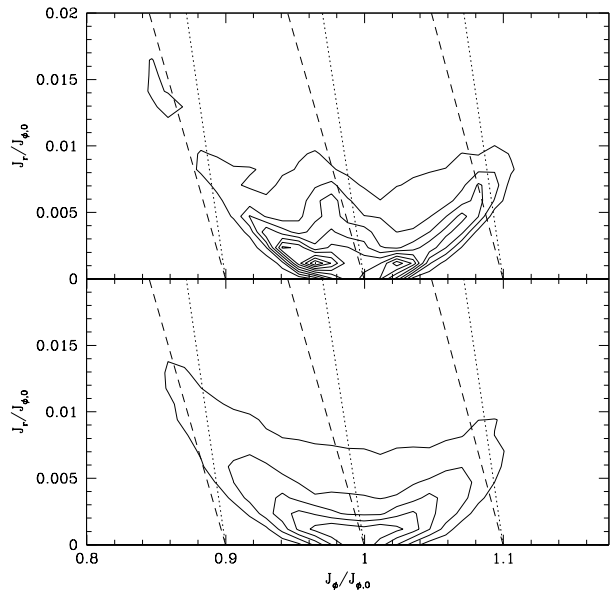}}}
  \caption{Contour plot of the density in the $J_r$, $J_\phi$ 
    plane of stars taken from the GCS data (\emph{solid} lines, 
    \emph{upper} plot), and of the phase-mixed model described in 
    Section~\ref{sec:num} (\emph{solid} lines, \emph{lower} plot). 
    Contours are spaced linearly in density, at the 
    same densities in each plot (relative to the total 
    number of stars in each case). 
    The Pleiades and Sirius moving groups produce small overdensities 
    around $(J_r,J_\phi) = 0.96,0.0015$ and $1.02,0.0015$ respectively 
    (in units where $J_{\phi,0}=1$). The Hyades moving group produces a
    overdensity which is spread out in $J_r$ at around 
    $J_\phi=0.97$, tending towards slightly lower $J_\phi$ with 
    increasing $J_r$.  Resonance lines for 2:1 OLRs (\emph{dotted}) and 
    ILRs (\emph{dashed}) are plotted, where the value of $\Omega_p$
    (eq.~\ref{eq:res_om}) is chosen in each case 
    such that the lines reach $J_r=0$ for $J_\phi=0.9,1$ and $1.1$. A 
    scatterplot of the distribution of the GCS stars in action was shown 
    by S10 as fig. 3.
\label{fig:actions}
}
\end{figure}

Thus far in this study I have only considered two of the conditions on 
resonant stars -- that they lie on or near lines of 
$l\theta_r+m\theta_\phi=const$ and that they must lie in the survey volume. 
To understand the expected distribution of 
resonant stars in a survey of the Solar neighbourhood, 
one must consider the combination of three different requirements on them:
\begin{itemize}
\item They must have angle coordinates near to a given resonance line in angle
  (eq.~\ref{eq:res}).
\item They must have orbital frequencies (and thus actions) which place them
  near to a given resonance line in frequency (eq.~\ref{eq:res_om}).
\item They must lie within the survey volume (e.g. within $200\pc$ of
  the Sun).
\end{itemize} 
Note that over the Galaxy as a whole, the requirement on $\bolth$ only 
affects the distribution in $\bolth$ and the requirement on $\bolJ$ only 
affects the distribution in $\bolJ$ -- the two distributions can be thought
of as independent\footnote{In practice they are not entirely independent -- the
resonant stars at the (say) higher values of $l\theta_r+m\theta_\phi$ about the 
resonance are probably there \emph{because} they have the higher values of
$l\Omega_r(\bolJ)+m\Omega_\phi(\bolJ)$. This effect is likely to have some 
impact, but is beyond the scope of this study}. 
It is only because of the finite survey volume, 
and therefore the finite range of $\bolth$ for which stars with  
for a given $\bolJ$ will be observed, that 
the $\bolJ$ condition significantly 
affects the observed $\bolth$ distribution (and vice versa).

\begin{figure*}
  \centerline{\hfil
    \resizebox{82mm}{!}{\includegraphics[angle=270]{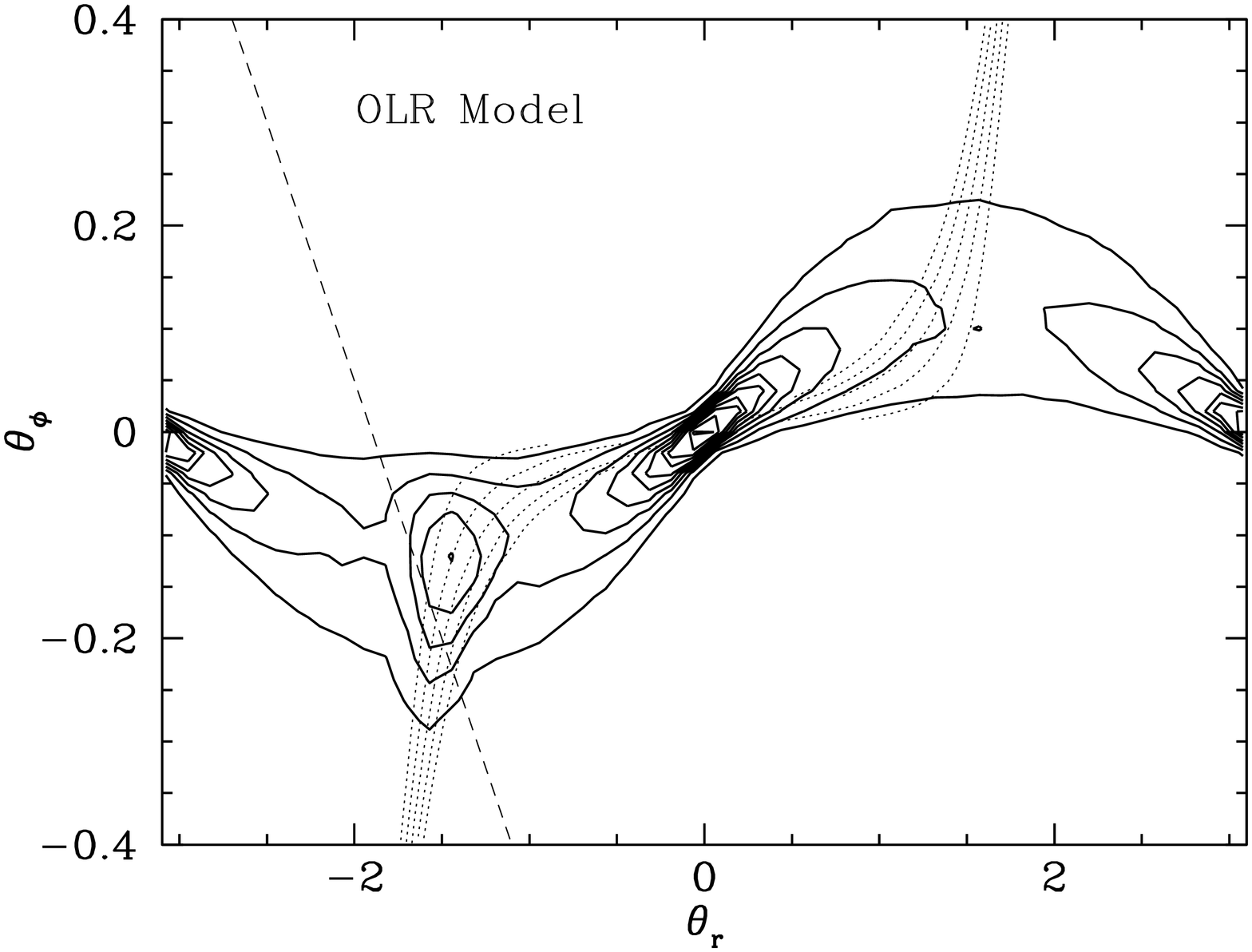}}
    \resizebox{82mm}{!}{\includegraphics[angle=270]{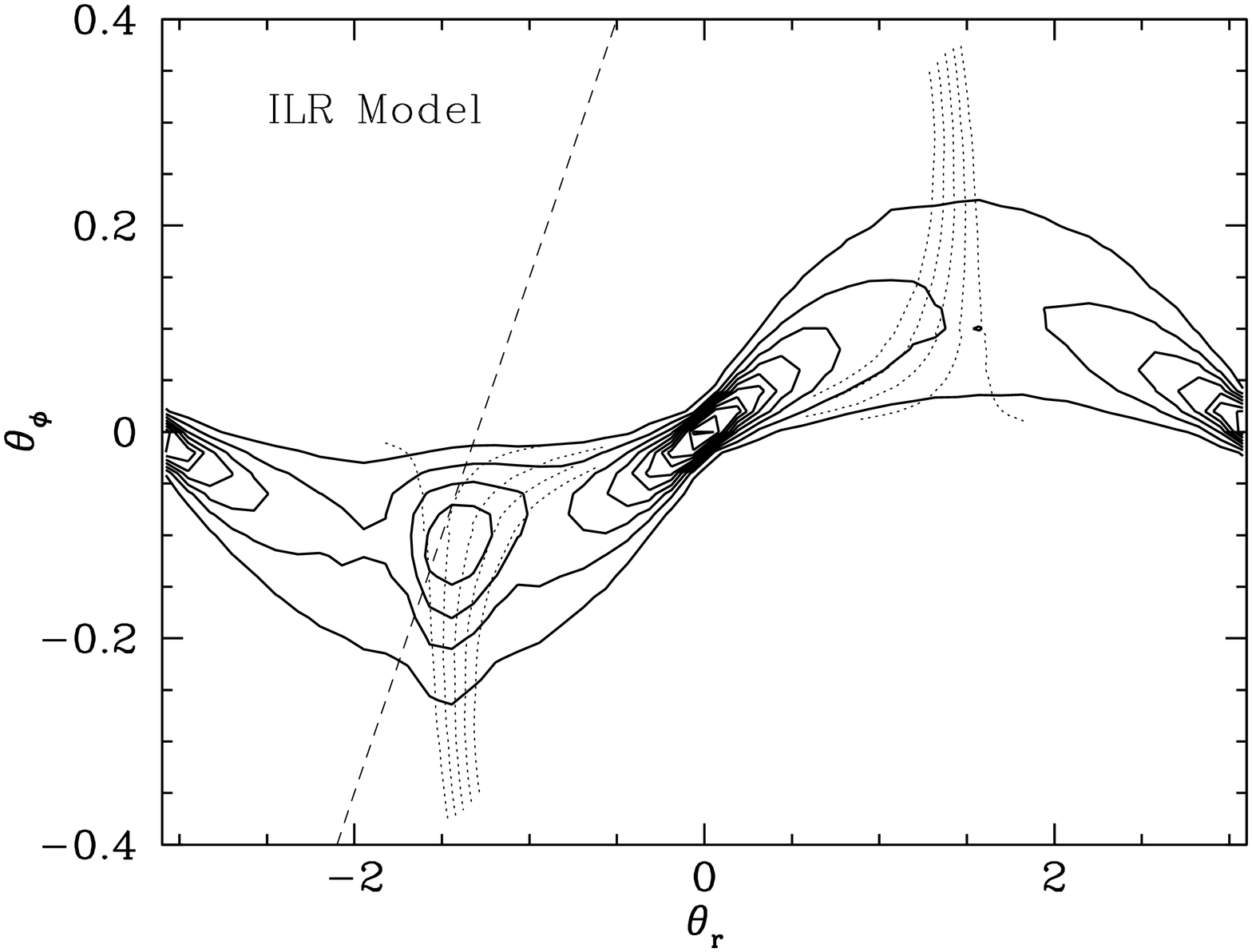}}} 
  \centerline{\hfil
    \resizebox{82mm}{!}{\includegraphics[angle=270]{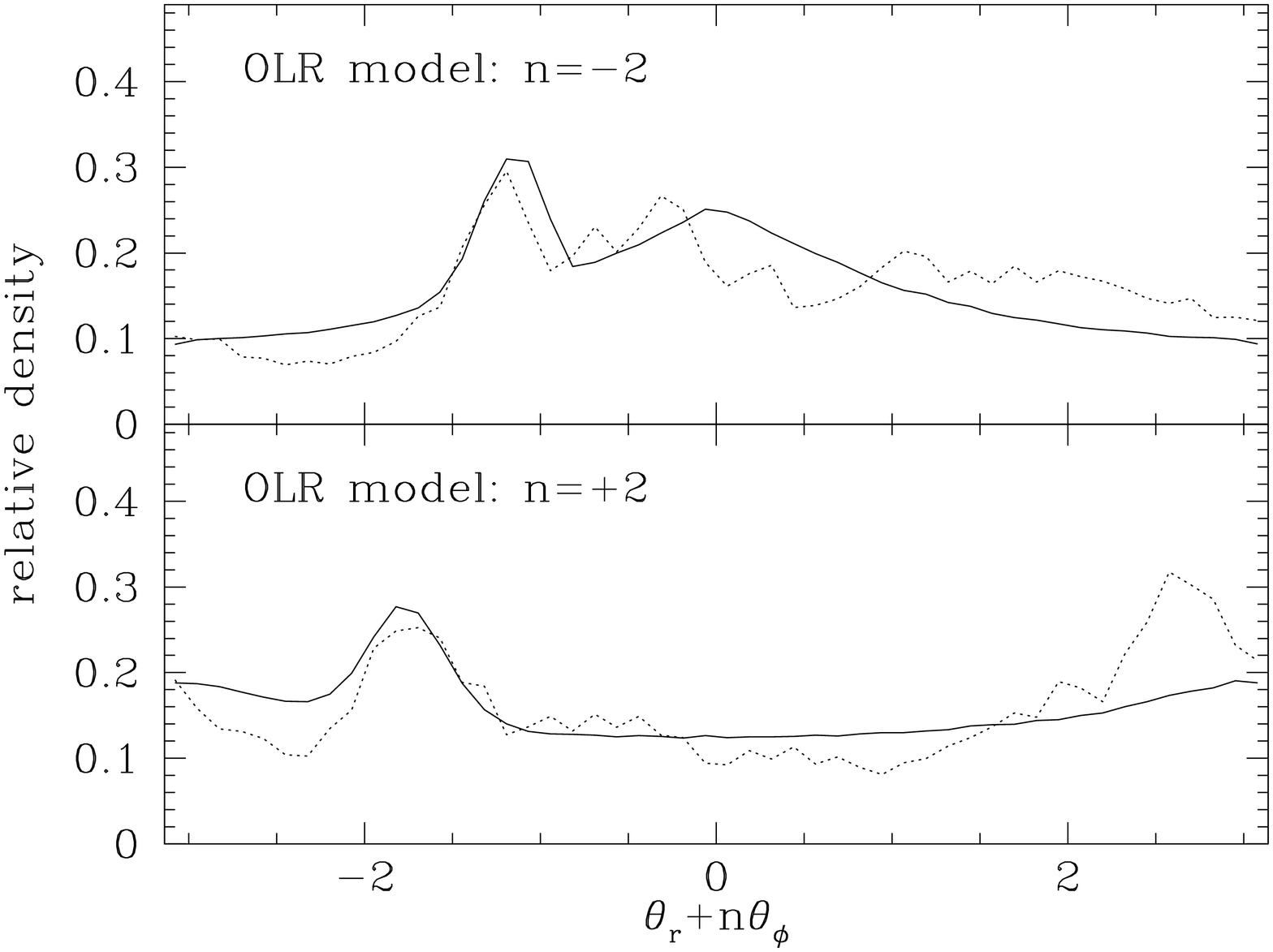}}
    \resizebox{82mm}{!}{\includegraphics[angle=270]{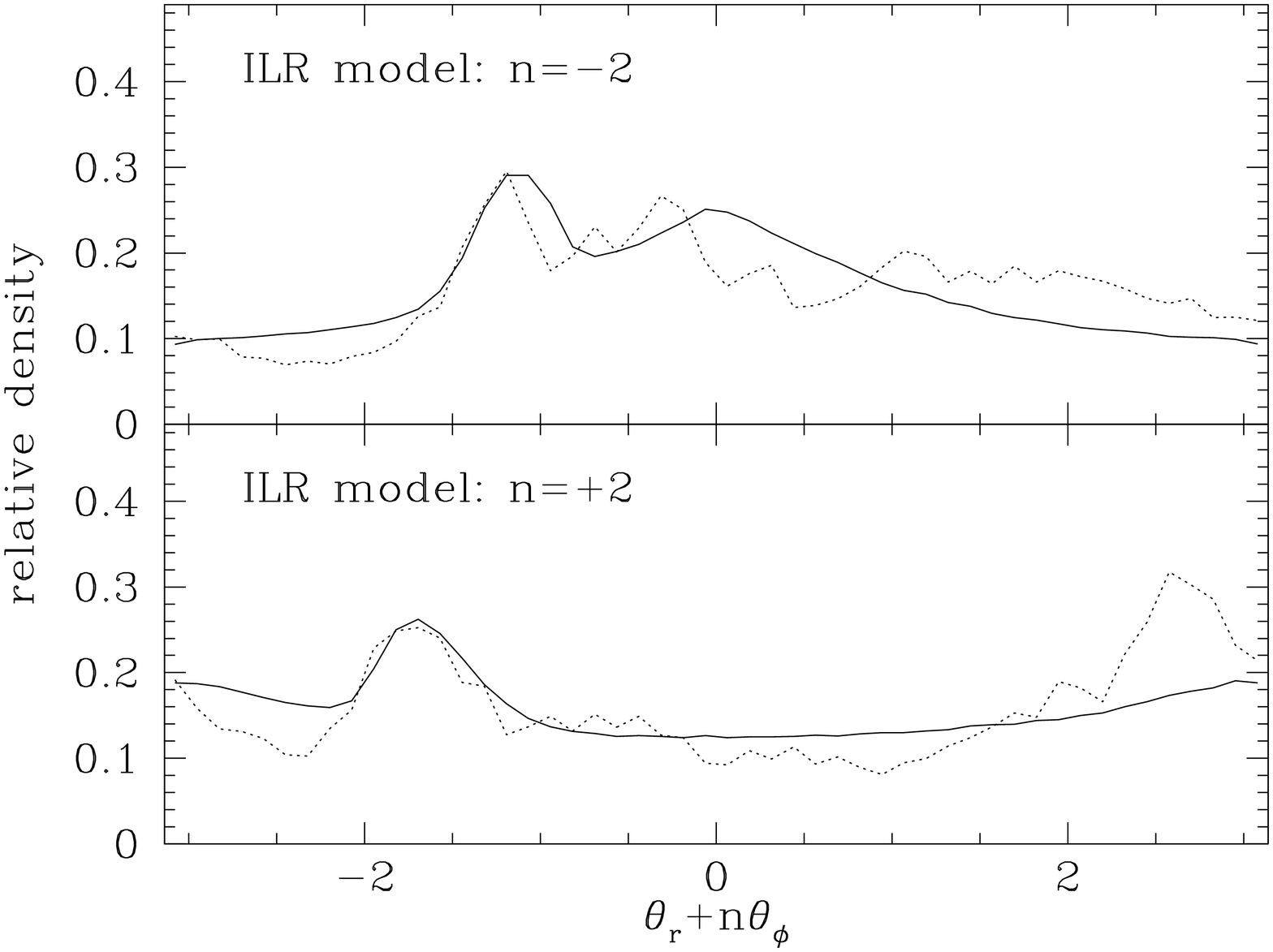}} }
  \caption{
    Plots showing the distribution of stars in $\bolth$ of stars taken 
    from the OLR (\emph{left}) and ILR (\emph{right}) models. The upper 
    panels show a contour plot of the density in the $\theta_r$, $\theta_\phi$ 
    plane (\emph{solid} line). The \emph{dashed} line in those figures is 
    the line $\theta_r=\theta_{r,res}(\theta_\phi)$ (eq.~\ref{eq:resline}), 
    and the \emph{dotted} lines are the angle coordinates corresponding to the
    resonance lines in action at the Sun's position  -- as discussed in 
    Section~\ref{sec:AAcoords} a position and value of $\bolJ$ defines
    two possible values of $\bolth$. In either set of five dotted lines, the 
    central one corresponds to $J_\phi=J_{\phi,res}(J_r)$, and those 
    either side correspond to $\pm\Delta_{J,res}$ and $\pm2\Delta_{J,res}$.
    The lower panels show the distribution of 
    stars in $\theta_r+n\theta_\phi$, as in Figure~\ref{fig:meat}
    except restricted to only $n=-2$ or $2$. In this case the 
    \emph{solid} line shows the distribution for the model, and the 
    \emph{dotted} line shows the distribution for the GCS stars 
    (for comparison).
   \label{fig:IOLR_mod_cont}
 }
\end{figure*}

In Figure~\ref{fig:actions} I plot the density of the GCS stars 
as a function of $J_\phi$ and $J_r$. The density of stars in my phase-mixed 
model is also plotted, for comparison. For a given $J_\phi$ there is 
a minimum $J_r=J_{r,min}$ for stars to reach the Solar neighbourhood, 
which can be thought of as a minimum epicyclic amplitude 
for a given guiding centre radius\footnote{For a given 
$J_\phi$, the value of $J_{r,min}$ is in fact dependent on $J_z$, with increasing
$J_z$ generally causing a decrease in $J_{r,min}$ if $J_\phi<J_{\phi,SN}$, 
and an increase in $J_{r,min}$ if $J_\phi>J_{\phi,SN}$, where $J_{\phi,SN}$ is
the angular momentum of a circular orbit that passes through the Solar
neighbourhood. This effect is small for the stars considered here as 
they have $J_z\ll J_\phi$.}. This is the cause of the near-parabolic
lower boundary seen in Figure~\ref{fig:actions}. The Pleiades and Sirius moving groups can be 
clearly seen as small overdensities in this plot. The Hyades moving group is 
seen as a rather more spread out overdensity at a range of $J_r$, around 
$J_\phi=0.97J_{\phi,0}$, tending towards slightly lower $J_\phi$ with 
increasing $J_r$.

The dotted and dashed lines in Figure~\ref{fig:actions}
are 2:1 OLR and ILR lines respectively, these are
lines along which $2\Omega_\phi(\bolJ)+\Omega_r(\bolJ)=2\Omega_p$ and
$2\Omega_\phi(\bolJ)-\Omega_r(\bolJ)=2\Omega_p$ respectively, for different
values of $\Omega_p$, the perturber pattern speed, chosen such that the 
resonance lines reach $J_r=0$ at $J_\phi=0.9,\,1$, or 
$1.1J_{\phi,0}$. Changing the value of 
$\Omega_p$ moves the resonance lines in $J_\phi$, but does not significantly
alter their gradient in this range of $\bolJ$. The Hyades overdensity seems 
to lie around a Lindblad resonance line, but this could be 
either an OLR or ILR line
-- it was this fact which lead S10 to claim this was an 
Lindblad resonance, but that one needed to investigate the distribution in 
angle to determine which one. Other resonances  -- the 3:1 or 4:1 OLR or ILR
lines -- would appear very similar on Figure~\ref{fig:actions}, 
though the 2:1 ILR line is the furthest from the vertical. It is also 
worth noting that the slope of the various resonance lines is sensitive to 
the Galactic potential -- in a logarithmic potential (of the kind used 
by S10), the gradients of the 2:1 OLR and ILR lines in this part of 
$\bolJ$-space are nearly identical. It may be possible to use the  
slope of resonance lines in action space to provide information about the
Galactic potential by comparing them to observed dynamical substructure, 
but that is beyond the scope of this study.

To explore the expected distribution of stars in the Solar 
neighbourhood associated with a resonance, I consider a \df \ related
to the phase-mixed \df \ used previously, adjusted to include
a resonant component:
\begin{equation} \label{eq:resdf}
f(\bolJ,\bolth) \propto f_0(\bolJ)\times(1+C\,\eta_{res}(\bolJ,\bolth))
\end{equation}
where $f_0$ is the distribution function described in Section~\ref{sec:num}, 
$C$ is a constant chosen such that the resonant component 
contributes $8$ percent of the stars observed in the Solar neighbourhood, and 
\begin{equation} \label{eq:resline}
  \eta_{res}(\bolJ,\bolth) = \exp\left(-\frac{(J_\phi-J_{\phi,res})^2} 
    {\Delta_{J,res}^2}-\frac{(\theta_r-\theta_{r,res})^2}
    {\Delta_{\theta,res}^2}\right).
\end{equation}
$J_{\phi,res}$ is a function of $J_r$ and is chosen such 
that $l\Omega_r(J_r,J_{\phi,res})+m\Omega_\phi(J_r,J_{\phi,res}) 
= const$, for $J_z=0$, and 
$\theta_{r,res}$ is a function of $\theta_\phi$ and is chosen such that  
$l\theta_{r,res}+m\theta_\phi = const$. The values 
$\Delta_{J,res}$ and $\Delta_{\theta,res}$ give the width of the resonance peak 
around the exact resonance lines in $J_\phi$ and $\theta_r$, respectively. 
One could, equally, describe the width in action or angle in terms of a 
spread in $J_r$ or $\theta_\phi$ respectively, but for convenience I have 
chosen to describe it in terms of the coordinates with the greater 
ranges of values in these data. The width $\Delta_{J,res}$ is effectively a 
width in frequency about the pattern speed of the perturber. In the toy models 
I show here I take $\Delta_{J,res}=0.01J_{\phi,0}$, $\Delta_{\theta,res}=0.3$. 

I consider two toy models, each designed to produce models with an 
overdensity in phase-space in a similar volume to that where the Hyades
moving group is found (but not tuned to produce a best fit), one corresponding 
to an OLR ($l=1$, $m=2$) 
and one corresponding to an ILR ($l=-1$, $m=2$). For the OLR model, I take 
$\theta_{r,res}+2\theta_\phi=-1.9$, and for the ILR model 
$-\theta_{r,res}+2\theta_\phi=1.3$. In the OLR case I take 
$J_{\phi,res}(J_r=0)=0.975J_{\phi,0}$, and in the ILR case 
$J_{\phi,res}(J_r=0)=0.985J_{\phi,0}$.

Figure~\ref{fig:IOLR_mod_cont}
shows contour plots of the density in the $\theta_r$, $\theta_\phi$ plane 
of the OLR and ILR models, and plots of $\theta_r+n\theta_\phi$ (as in 
Figure~\ref{fig:meat}) restricted to $n=\pm2$ in the interests of brevity. Both
the ILR and OLR models reproduce some of the features of the Hyades 
overdensity. In both cases 
the overdensity in angle space is somewhat triangular in shape, like the Hyades
overdensity, rather than following a single line as one would 
expect if only the condition on angle (eq.~\ref{eq:res}) was relevant. In both 
cases the overdensity in angle is strong for the two cases 
$n=\pm2$, as well for other values of $n$ (not shown). 

In an effort to explain the 
structure of the overdensity in the $\theta_r$, $\theta_\phi$ plane, 
the upper panels of 
Figure~\ref{fig:IOLR_mod_cont} also show the lines $\theta_r=\theta_{r,res}$
for the two models, and lines corresponding to the condition on $\bolJ$. The 
latter are found by taking the condition that $J_\phi=J_{\phi,res}(J_r)$ (or
$J_\phi=J_{\phi,res}\pm\Delta_{J,res}$ or $J_\phi=J_{\phi,res}\pm2\Delta_{J,res}$)
and determining the two possible 
values of $\bolth$ that a star with these actions 
would have at the Sun's position -- in the relevant part of phase space, 
lower values of $J_r$ correspond to smaller (i.e. closer to zero) values 
of $\theta_\phi$. This gives a sense of the two competing 
effects which (in addition to the general selection effects illustrated in
Figure~\ref{fig:mod_0_cont}) determine the shape of the overdensity in 
angle space, but it is important to also consider the effect of the finite
volume surveyed on the constraint applied by the condition on $\bolJ$, 
i.e. how close to the dotted lines in Figure~\ref{fig:IOLR_mod_cont} stars 
must actually lie. 

A star with a given value of $\bolJ$ which lies 
within $200\pc$ of the Sun will have a value $\theta_\phi$ which 
lies within $\sim0.02$ of the value for a star \emph{at} the Sun, while the
range of possible values of $\theta_r$ is typically larger, and increases 
with decreasing $J_r$ (and thus, in the overdensity considered, with 
decreasing $\theta_\phi$). This latter point can be understood by 
considering a star in the epicycle approximation. As 
$J_r\rightarrow0$ the epicycle shrinks to negligible size, so the 
star can be at any point on the epicycle (therefore any $\theta_r$) and still 
be within the survey volume if the value of $\theta_\phi$ is one that 
places it there. As $J_r$ increases, and the epicycle increases in size, 
the fraction of the epicycle (i.e. the range in $\theta_r$) that 
corresponds to stars that enter the survey volume decreases.

\begin{figure}
  \centerline{\resizebox{\hsize}{!}{\includegraphics{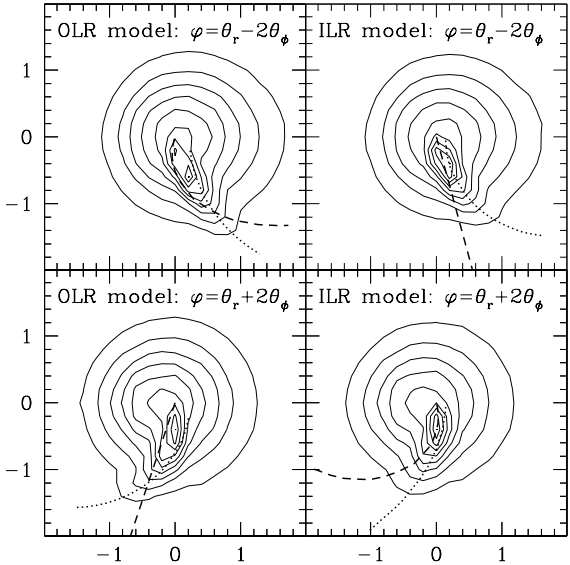}}}
  \caption{Contour plot showing the distribution of stars from the OLR
    (\emph{left}) and ILR (\emph{right}) models in polar 
    coordinates with radial coordinate $a$, 
    the amplitude of radial motion in the 
    epicycle approximation, and polar coordinate 
    $\varphi=\theta_r+n\theta_\phi$ with $n=-2$ (\emph{upper}) 
    or $n=2$ (\emph{lower}). The \emph{dashed} lines are given by 
    $\theta_r=\theta_{r,res}(\theta_\phi)$ (eq.~\ref{eq:resline}) for the 
    given model, evaluated at the Sun's position (so that the value of $\bolth$
    determines a value of $\bolJ$). The \emph{dotted} lines are given by 
    $J_\phi=J_\phi(J_{r,res})$ for the given model, again evaluated at 
    the Sun's position. 
\label{fig:IOLR_mod_pol}
}
\end{figure}

Figure~\ref{fig:IOLR_mod_pol} shows polar plots, 
analogous to those in Figure~\ref{fig:pol_GCS}, each labeled with 
which model they represent and what was used as the polar angle 
$\varphi$ (the radial coordinate, as before, is always radial epicycle
amplitude $a$). 
In each case I again plot the lines corresponding to 
$\theta_r=\theta_{r,res}(\theta_\phi)$ and $J_\phi=J_{\phi,res}(J_r)$, evaluated at
the Sun's position. Once again, it is clear that both resonant conditions 
are important on the distribution in these plots, so the overdensity seen
will be distorted away from $\theta_r+n\theta_\phi=const$ 
even if the model has $\theta_{r,res}+n\theta_\phi=const$ as a constraint, 
providing there is a non-negligible width around the resonance in $\bolth$.

Note that as $J_r$ increases (and the size of the epicycle increases), 
the condition on $\bolJ$ becomes more 
restrictive in $\bolth$ for the reason discussed above. Therefore this 
condition becomes increasingly dominant on the observed overdensity. 
This can be seen in Figure~\ref{fig:IOLR_mod_pol} as the outer contours show 
that the overdensity appears to follow the dotted line
corresponding to $J_\phi=J_{\phi,res}(J_r)$ at large $J_r$. For \emph{both} of 
the models this means that the contours seem to show an overdensity that 
lies around a value of $\theta_r-2\theta_\phi$ that is approximately 
constant with increasing $J_r$, while lying around a value of 
$\theta_r+2\theta_\phi$ that varies with increasing $J_r$. This provides a 
natural explanation for the slight curvature as $J_r$ increases of the 
overdensity associated with the Hyades moving group in the equivalent 
plot for the GCS stars (Figure~\ref{fig:pol_GCS}) -- for overdensities
produced by either an OLR or ILR, 
this curvature is seen as a result of the selection effects 
and the constraint in $\bolJ$.

I have explored models with resonances at other frequency ratios (and 
thus with different relationships between $\theta_r$ and $\theta_\phi$), and
found that it is possible to produce overdensities in $\bolth$ that are 
qualitatively very similar to those shown here. While it may be possible
to tell one from another for given observations (such as those of the 
Hyades) it is certainly a very complicated task, and one that will
require careful modelling and analysis.


\section{Discussion and conclusions} \label{sec:conc} 
In this paper I have re-examined the distribution of stars in the 
Solar neighbourhood in angle coordinates following 
the claim by S10 that the Hyades moving group is related to an ILR.
Using a dynamical ``torus'' model I showed the significant impact
of selection effects associated with surveying a finite (small) volume upon the
distribution of stars in angle coordinates taken from a phase-mixed model.
Using models which contain resonant components in addition to a 
phase-mixed background I have demonstrated the important effects that the 
distribution of resonant stars in action have on the observed distribution 
in angle (again because of selection effects).

The distribution of the stars associated with the Hyades moving group
in action (Figure~\ref{fig:actions}) indicates that it 
is associated with some resonance between the radial oscillations and
the azimuthal motion (i.e. follows a relation of the form 
of $l\Omega_r+m\Omega_\phi$). It is also clear that the stars which make up 
the resonant component are constrained in $\bolth$ in some way, otherwise 
there would be an overdensity at $\theta_r\sim1.5$ as well as 
$\theta_r\sim-1.5$. However, it is very difficult to determine 
what the values of $l$ and $m$ are, and thus which resonance is responsible. 
In action 
space the lines $l\Omega_r(\bolJ)+m\Omega_\phi(\bolJ)=const$ are very 
similar, in the relevant range of $\bolJ$, for different values of $l$ and 
$m$, and are sensitive to choice of the Galactic potential. The 
approach taken by S10, of looking for overdensities in the statistic
$l\theta_r+m\theta_\phi$, is also flawed because of the selection 
effects that have important influences on the distribution in $\bolth$. 
I find that, contrary to the conclusions of S10, it not clear that 
the Hyades moving group is the result of an inner Lindblad resonance. 

It may be possible to use careful modeling and analysis, which takes 
into account selection effects and the 
distributions in $\bolth$ and $\bolJ$ simultaneously, to determine 
what type of resonance is responsible for the Hyades moving group from these
data. However this task is made even more difficult by uncertainty about 
the Galactic potential which significantly affects the gradient of resonance
lines in action space.

It seems likely the problems of selection effects in studies of this 
kind will prove extremely challenging, if not intractable, while we are
dealing with survey volumes corresponding to the relatively small
radii ($\sim200\pc$) associated with the GCS. To go beyond this with
similar accuracy (i.e.~$\sim1\kms$ velocity uncertainty) 
requires distance and proper measurements more accurate than
those currently available -- the RAVE and SDSS data used by 
\cite{HaSePr11} has significantly larger uncertainties even though they
select stars which lie in the same volume as the GCS stars. 
Gaia \citep{GAIA01} is expected to produce
3-dimensional velocity measurements accurate to $\sim1\kms$ for some
stars up to $\sim3\kpc$ from the Sun, corresponding to a range in
Galactocentric $\phi$ of order $0.7$ radians. This will dramatically 
reduce the impact of selection effects on the observed distributions in
angle-action coordinates, and should make it relatively straightforward to 
pick out structures in angle space that are unambiguously of the kind 
predicted by S10. This process is likely to both be guided by and 
affect the determination of the gravitational potential of the Galaxy 
(an essential product of the Gaia mission).

Throughout this study I have assumed that the uncertainties on the positions 
and velocities of the stars observed by the GCS can be ignored, because they 
are too small to have any significant effect. It is worth noting that the 
relationship between the uncertainty in $\bolx,\bolv$ and that in 
$\bolth,\bolJ$ is not entirely straightforward, and that a small uncertainty
in velocity does not necessarily correspond to a small uncertainty in 
$\bolth$. As inspection of Figure~\ref{fig:UVaxes} suggests, the 
relationship between an uncertainty in velocity and that in $\theta_r$ is 
heavily dependent on $J_r$, with smaller $J_r$ corresponding to a 
greater uncertainty in $\theta_r$ for given uncertainty in $\bolv$.
Indeed as $J_r\rightarrow0$ a negligible uncertainty in $\bolv$ 
corresponds to an uncertainty of $2\pi$ in $\theta_r$. The 
Hyades overdensity does not extend to $J_r=0$ (as can be seen in 
Figure~\ref{fig:UV_GCS}), so the assumption that the uncertainty in
$\theta_r$ can be ignored is reasonable for this study. However, 
future efforts to understand resonances of this type is likely to 
require appropriate treatment of these uncertainties.


\section*{Acknowledgments}
I thank members of the Oxford dynamics group for helpful input, 
especially James Binney for careful reading of an early 
draft of this paper. I am also grateful to Jerry Sellwood for a very 
helpful discussion which provoked the work described in 
Section~\ref{sec:resmod}.  This work is supported by a grant from the 
Science and Technology Facilities Council.

\bibliographystyle{mn2e} \bibliography{new_refs}


\end{document}